\documentstyle[epsf]{preprint}
\journalid{*}{*}{1998}
\articleid{1}{5}
\newcommand{\bc}{\begin{center}}
\newcommand{\ec}{\end{center}}

\def\ltsima{$\; \buildrel < \over \sim \;$}
\def\ltsim{\lower.5ex\hbox{\ltsima}}
\def\gtsima{$\; \buildrel > \over \sim \;$}
\def\gtsim{\lower.5ex\hbox{\gtsima}}

\title{NEW INSIGHTS INTO THE EARLY STAGE OF THE GALACTIC CHEMICAL EVOLUTION}
{THE EARLY STAGE OF THE GALACTIC CHEMICAL EVOLUTION}
\author{
 Takuji Tsujimoto \\
{\em National Astronomical Observatory, Mitaka, Tokyo, 181-8588 Japan;} \\
{\em tsuji@misty.mtk.nao.ac.jp} \\
Toshikazu Shigeyama\\
{\em Department of Astronomy, School
of Science, University of Tokyo, Bunkyo-ku, Tokyo, 113-0033 Japan}\\
{\em Research Center for the Early Universe, School of Science, University
of Tokyo,  Bunkyo-ku, Tokyo, 113-0033 Japan;} \\
{\em shigeyama@astron.s.u-tokyo.ac.jp}}
{T. Tsujimoto \& T. Shigeyama}

\received{10 September 1998}
\accepted{1 October 1998}

\abstract{
The supernova yields of several heavy elements including $\alpha$-,
iron-group, and r-process elements are obtained as a function of the
mass of their progenitor main-sequence stars $M_{\rm ms}$ from the
abundance patterns of extremely metal-poor stars with a procedure
recently proposed by Shigeyama and Tsujimoto (1998). The ejected
masses of $\alpha$- and iron-group elements increase with $M_{\rm
ms}$, whereas more Eu is ejected from supernovae with lower $M_{\rm
ms}$. For these several heavy elements, it is shown that the average
abundance ratios weighted by the Salpeter initial mass function
coincide with the ratios observed in stars with $-2<[{\rm Fe/H}]<-1$
within 0.1 dex. It follows that the correlations of stellar abundance
ratios with the metallicity are twofold. One is the abundance ratios
for [Fe/H] $ <-2.5$ imprinted by the nucleosynthesis in individual
supernovae on the timescale $\sim 10^7$yr and the other for [Fe/H]
$>-2$ results from the mixing of the products from a whole site of the
nucleosynthesis, taking place on the timescale longer than $10^9$ yr.
}


\begin{document}
\section{INTRODUCTION}

Older stars in the Galactic halo show poorer abundances of heavy
elements in their spectra if most of heavy elements are well mixed in
the halo after they are synthesized in massive stars and ejected by
type II supernova (SN II) explosions. Shigeyama \& Tsujimoto (1998)
have shown that this age-metallicity relation is broken for extremely
metal-poor stars with [Fe/H] $<$ --2.5 recently observed (McWilliam et
al. 1995; Ryan, Norris, \& Beers 1996). Instead their spectra show
elemental patterns of individual first generation SNe. This must be a
result from that most of the stars with [Fe/H]$<-2.5$ were formed from
individual supernova remnants (SNRs). It means that utilizing the
observed abundance patterns of extremely metal-poor stars, one can
construct yields of any observed heavy-element from each SN II.

Shigeyama \& Tsujimoto (1998) have also shown that the mass of Fe
ejected by each SN II ($M_{\rm Fe}$) as a function of progenitor 
mass at the main sequence ($M_{\rm ms}$) can be derived from the
observed [Mg/Fe]$-$[Mg/H] trend combined with the [Mg/H]$-M_{\rm ms}$
relation in theoretical SN models and that this $M_{\rm Fe}$ as a
function of $M_{\rm ms}$ is consistent with that derived from SN light
curve analyses. However this kind of check cannot be applied to the
other elements. In this {\it letter}, we examine whether the yields of
various elements other than Fe obtained by this procedure can
reproduce the observed plateau of [element/Fe] seen in the halo stars
for $-2<$[Fe/H]$<-1$, which results from a whole population of SNe II
integrated over the initial mass function (IMF).

Although the $r$-process nucleosynthesis is poorly understood (Mathews
\& Cowan 1990), our scenario can extract information of the site for
production from the observed abundances of $r$-process elements on the
surface of metal-poor stars. At the same time, it provides a new
picture of the chemical evolution for $r$-process elements. Europium
is a good indicator of the $r$-process because it is about 90\% of
pure $r$-process origin in the Solar System (Howard et al. 1986) and
its abundances in stars have been observed by many authors (Butcher
1975; Luck \& Bond 1985; Gilroy et al. 1988; Magain 1989; Woolf,
Tomkin, \& Lambert 1995 McWilliam et al. 1995; Ryan et al. 1996). The
observed trend of [Eu/Fe]-[Fe/H] for [Fe/H]$<-2$ before McWilliam et
al. (1995) and Ryan et al. (1996) was regarded as the decrease in the
[Eu/Fe] ratio as [Fe/H] decreases, which was ascribed to the later
production of the $r$-process element than that of iron; The
$r$-process occurs in low-mass SNe II with initial masses of $10-11
M_{\odot}$ (Mathews \& Cowan 1990; Mathews, Bazan, \& Cowan 1992). The
observations (McWilliam et al. 1995; Ryan et al. 1996) covering stars
with [Fe/H] $\sim -4$ have drastically changed the observed trend of
[Eu/Fe]-[Fe/H] for [Fe/H] $< - 2$, indicating the increase in the
[Eu/Fe] ratio as [Fe/H] decreases. Such an observed trend is
incompatible with the possible $r$-process sites proposed so far,
whereas from our point of view, the observed trend in the range of
[Fe/H] $< -2.5$ reflects different [Eu/Fe] ratios in SNRs with
different progenitor masses.

In this {\it letter}, we will show that the heavy-element yields for
$\alpha$-, iron-group elements and Eu derived from the analysis of
abundance pattern in extremely metal-poor stars are quite reliable
ones and that combining yields thus obtained with the Galactic
chemical evolution model, the observed abundance patterns over the
whole metallicity range can be well reproduced.

\section{WELL-MIXED OR INHOMOGENEOUS?}

The yields of heavy elements not influenced by the mass cut between
the forming neutron star (or black hole) and the ejected envelope are
expected to be more reliably calculated in SN II models than those
influenced by the mass cut. Therefore the observed abundance ratios of
[C/Mg] with respect to [Mg/H] can tell whether extremely
metal-poor stars were formed from individual SNe or from the
well-mixed interstellar gas. The observed correlation of [C/Mg] with
[Mg/H] (McWilliam et al. 1995) is shown in Figure 1. The dashed curve
in the same figure is the correlation of these ratios predicted from
the one-zone Galactic chemical evolution model in which the gas is
distributed uniformly and the heavy elements are well-mixed within the
zone. Here we use the updated nucleosynthesis calculations by Nomoto
et al. (1997) for SN II yields and the results of van den Hoek \&
Groenewegen (1997) for the AGB star's yields. Details of the basic
model ingredients are presented in Yoshii, Tsujimoto, \& Kawara
(1998). This obvious disagreement with the observation indicates that
the mixing of elements did not take place at that time.

\begin{figure}[ht]
\begin{center}
\leavevmode
\epsfxsize=\columnwidth\epsfbox{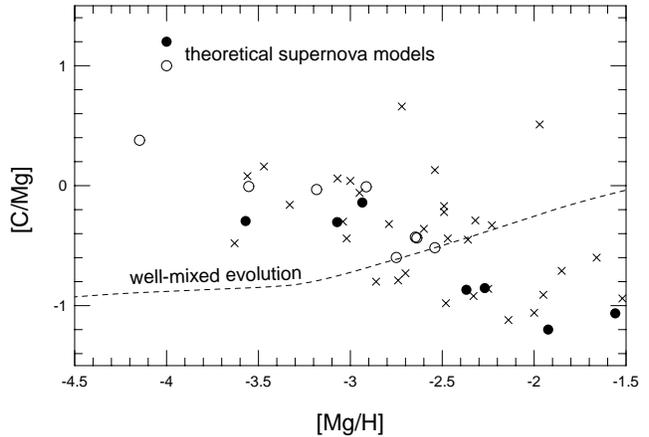}\hfil
\end{center}
\caption{Comparison of the observed correlation of [C/Mg] -- [Mg/H]
(crosses: McWilliam et al. 1995) with the theoretical nucleosynthesis
[C/Mg] ratio from type II supernovae as a function of the predicted
[Mg/H] (filled circles: Nomoto et al. 1997; open circles: Woosley \&
Weaver 1995). The evolutionary change in [C/Mg] predicted by the
one-zone chemical evolution model including the contribution from AGB
stars (van den Hoek \& Groenewegen 1997) is also shown by the dashed
curve.}
\end{figure}

The open and filled circles are the correlation of the abundance
ratios in SNRs at the end of expansions calculated from two
theoretical SN models (Nomoto et al. 1997; Woosley \& Weaver
1995). The coincidence of the two correlations must be a result from
that most of the stars with [Mg/H]$<-2$ were formed from individual
SNRs. Therefore the observed correlation must be made within the first
$\sim2\times 10^7$ yr (corresponding to the life time of the least
massive star that explodes as a SN II) from the birth of the first
generation stars. During this period, a SNR scarcely intersects with
another SNR (Audouze \& Silk 1995).

\section{THE IMF-WEIGHTED ABUNDANCE RATIOS}

The observed correlations of abundance ratios in halo stars with
higher [Fe/H] ([Mg/H]) result from the mixing of elements in the
Galactic halo. How precise the abundance pattern of stars for the
range of $-4<$[Fe/H]$<-2.5$ represents the nucleosynthesis yield of
each supernova can be checked by the comparison of their average
nucleosynthesis yield with the observed plateau (e.g., Gilroy et
al. 1988; Gratton 1989; Magain 1989; Gratton \& Sneden 1991; Nissen et
al. 1994) seen in the [element/Fe] ratio for $-2<$[Fe/H]$<-1$, which
results from a whole population of SNe II integrated over the
IMF. Using the same method as applied to the iron yield, we can obtain
the supernova yields for $\alpha$- and iron-group elements as a
function of the initial stellar mass. The procedure is as follows;

\begin{enumerate}
\item For each SN, the [Mg/H] ratio is calculated from the ratio of
the mass of Mg ejected from a SN II to the mass of hydrogen swept up
by the SNR (Shigeyama \& Tsujimoto 1998) and thus the theoretical
[Mg/H]-$M_{\rm ms}$ relation is derived.
\item We perform the $\chi ^2$ fit of the observational data in the
[Mg/element]-[Mg/H] diagram. 
\item We combine this curve with the theoretical [Mg/H]-$M_{\rm ms}$
relation and obtain the [Mg/element] ratio for each $M_{\rm ms}$.
\item Using the theoretical nucleosynthesis mass of Mg for each $M_{\rm ms}$, 
the ejected mass of element as a function of $M_{\rm ms}$ is derived.
\end{enumerate}

\begin{table*}
\caption{The comparison of the predicted abundance ratios of the halo 
stars in the range of $-2<$[Fe/H]$<-1$ with the observations}
\begin{tabular}{ccccccc} \hline \hline
\footnotesize
&\multicolumn{1}{c}{prediction} & & 
\multicolumn{4}{c}{observation} \\ 
\cline{3-7}
[ele/Fe] &  &  Nissen(94) & 
Gratton\&Sneden(91) & Gratton(89) & 
Magain(89) & Gilroy(88) \\ \hline
{[Mg/Fe]} & $+0.43$  & $+0.41\pm0.07$ & --- & --- 
& $+0.47\pm0.09$ & --- \\ 
{[Ca/Fe]} & $+0.44$  & $+0.35\pm0.06$ & $+0.29\pm0.06$ & --- 
& $+0.47\pm0.08$ & $+0.25\pm0.20$ \\ 
{[Cr/Fe]} & $-0.19$  & $-0.11\pm0.07$ & $-0.04\pm0.05$
& --- & $+0.01\pm0.08$ & $-0.12\pm0.24$ \\ 
{[Mn/Fe]} & $-0.42$  & --- & $-0.31\pm0.07$
& $-0.34\pm0.06$ & --- & --- \\ 
{[Co/Fe]} & $+0.02$  & --- &$-0.12\pm0.05$ & --- & --- & $-0.37\pm0.34$
\\ 
{[Ti/Fe]} & $+0.36$  & $+0.27\pm0.06$ & $+0.28\pm0.10$
& --- & $+0.40\pm0.09$ & $+0.08\pm0.10$ \\ 
{[Ni/Fe]} & $+0.09$  & --- & $-0.04\pm0.08$ & --- & --- & 
$-0.08\pm0.19$ \\ 
\hline
\end{tabular}
\end{table*}

\begin{figure}[ht]
\begin{center}
\leavevmode
\epsfxsize=\columnwidth\epsfbox{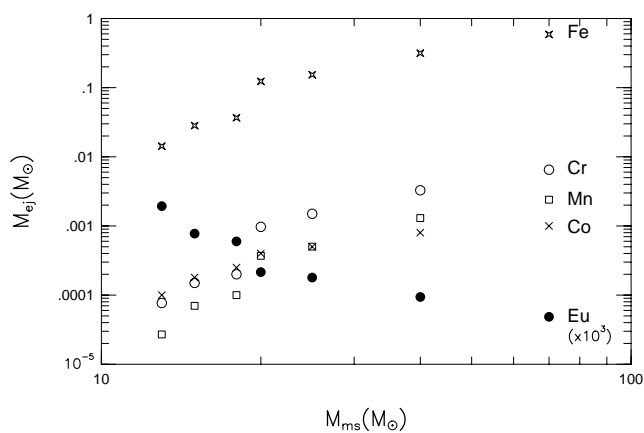}\hfil
\end{center}
\caption{Ejected masses of heavy elements from SNe II as a function
of the mass of their progenitor main-sequence stars. They are derived
from the observed abundance pattern in extremely metal-poor stars
combined with the [Mg/H]-$M_{\rm ms}$ relation in the theoretical SN
models.}
\end{figure}

The yields thus obtained are shown for several elements in Figure
2. Here we use the theoretical supernova model by Nomoto et
al. (1997). The iron-group elements are produced in the deepest layers
of the ejecta. Therefore a slightly different stellar-mass dependence
of the Co yield from other iron-group elements may indicate how the
explosion separates material comprising the neutron star from that of
the ejecta. We then calculate the average nucleosynthesis yields
integrated over $10 M_\odot< M_{\rm ms}< 50 M_\odot$ assuming the
Salpeter IMF, and obtain the average [element/Fe] ratios. They are
tabulated with the observed values in Table 1. The predicted average
abundance ratios are all consistent with the observed ones within an
error of $\sim 0.1$ dex. This good agreement strongly supports the
rightness of our determination of each supernova yield and suggests
that the observed correlations of stellar abundance ratios with the
metallicity are twofold: the nucleosynthesis abundance ratios for
[Fe/H] $< -2.5$, which in each star are imprinted by the
nucleosynthesis of a single supernova and the average abundance ratios
for [Fe/H] $> -2$, which result from a whole site for production. 

\begin{figure}[ht]
\begin{center}
\leavevmode
\epsfxsize=\columnwidth\epsfbox{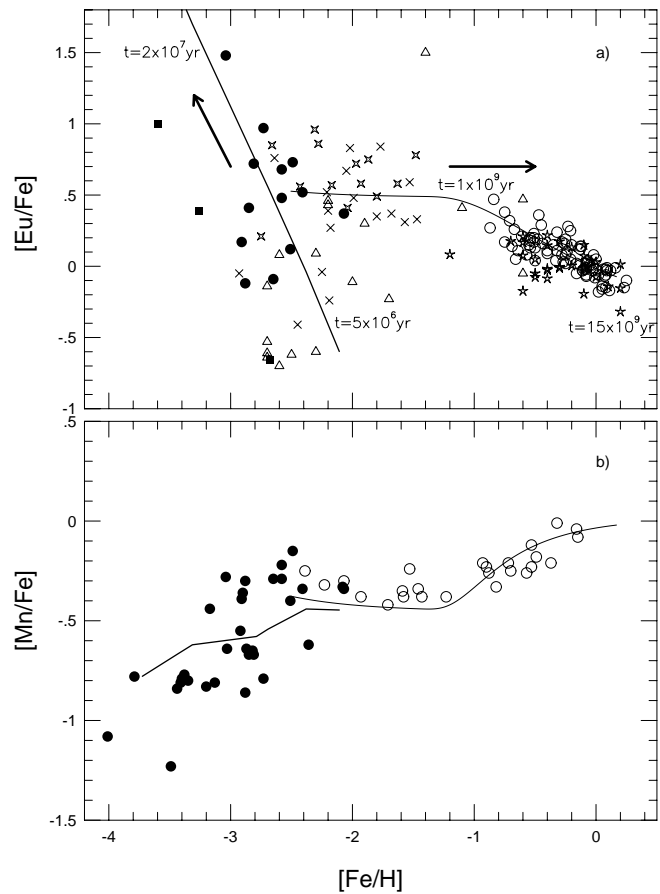}\hfil
\end{center}
\caption{(a) Correlation of [Eu/Fe] with [Fe/H]. The symbols
represent the data taken from various refs (filled circles: McWilliam
et al. 1995; filled squares: Ryan et al. 1996; open triangles: Luck \&
Bond 1985; four-pointed stars: Gilroy et al. 1988; crosses: Magain
1989; open circles: Woolf et al. 1995; five-pointed stars: Butcher et
al. 1975). See text for details. (b) the same as (a) but for
[Mn/Fe]. The circles are data from McWilliam et al. (1995) (filled
circles) and Gratton (1989) (open circles). See text for details.}
\end{figure}

It
is noted that the present nucleosynthesis models (Nomoto et al. 1997;
Woosley \& Weaver 1995) cannot reproduce the observed plateau values
for most of the elements in Table 1 due to a problem in modelling a SN
II explosion or mass cut (Thielemann, Hashimoto, \& Nomoto
1990). There are indications from Galactic evolution models (Timmes,
Woosley, \& Weaver 1995; Samland 1998) that the nucleosynthesis models
overestimate the Fe yields by a factor of two. However our results give
almost the same average Fe yield as those predicted by the
nucleosynthesis models and suggest that the heavy-element yields
predicted by them should be revised except for a few elements such as
C, Mg and O.

\section{$R$-PROCESS ELEMENT}

Figure 3a shows the observed correlation of [Eu/Fe] as a function of
[Fe/H]. The filled symbols (McWilliam et al. 1995; Ryan et al. 1996)
indicate the increase in the [Eu/Fe] ratio as [Fe/H] decreases for the
range of [Fe/H] $< - 2$, which  can be ascribed to different
[Eu/Fe] ratios in SNRs with different progenitor masses.

The ejected mass of Eu as a function of the initial stellar mass can
be obtained by the same method as for $\alpha$- and iron-group
elements (see \S 3), i.e., derived from the observed [Eu/Mg]-[Mg/H]
trend (McWilliam et al. 1995) combined with the [Mg/H]-$M_{\rm ms}$
relation in the theoretical SN models, and is shown in Figure
2. What's interesting is that Eu is ejected more in a less massive
star, different from the cases for other elements. This implies that
Eu is produced more in less massive stars or that more Eu eventually
falls back onto the nascent neutron stars in massive SNe II. The thick
line in Figure 3a indicates the predicted nucleosynthesis [Eu/Fe]
ratios in $18-50 M_{\odot}$ stars. The lower mass limit corresponds to
the most metal-deficient star observed by McWilliam et al. (1995).  As
a more massive star's explosion corresponds to a higher [Fe/H], the
time elapses toward lower metallicity as indicated by the arrow for
[Fe/H] $< -2.5$ in Figure 3a.

Using the nucleosynthesis yields of Fe and Eu inferred from extremely
metal-poor stars, we calculate the time evolution of [Eu/Fe] with the
one-zone model of Galactic chemical evolution (Yoshii et al. 1998) and
the result is shown in the [Eu/Fe]-[Fe/H] diagram for [Fe/H] $>-2.5$
(the thin line in Figure 3a). The predicted plateau in
$-2<$[Fe/H]$<-1$ is consistent with the observed ratios. The decrease
in [Eu/Fe] as [Fe/H] increases from $-1$ to $0$ is caused by the iron
production from type Ia supernovae, which eject their iron $\sim 1$Gyr
later. The sequence of the predicted nucleosynthesis [Eu/Fe] ratio for
[Fe/H] \ltsim $-2$ (thick line) does not connect smoothly with the
[Eu/Fe] plateau. It is a consequence of the decreasing mass of Eu with
the increasing stellar mass as shown in Figure 2, which leads to the
relatively small contribution from the massive SNe II to the
IMF-averaged value. This gap might yield the large scatter in [Eu/Fe]
until the [Eu/Fe] ratios converge to the average plateau ratio. As
shown by the two solid lines in Figure 3a, the observed abundance
ratios are created in two different enrichment epochs with different
time scales; the sequence of the first SN II explosions and the phase
approximated by the well-mixed evolution over several Gyrs in which
the age-metallicity relation holds. For comparison, we show that the
evolutionary change in [Mn/Fe] against [Fe/H] is also composed of two
different phases (Figure 3b). The [Mn/Fe] ratio from individual SNe II
at [Fe/H] $\sim -2.5$ connects with the [Mn/Fe] plateau, because the
IMF-averaged [Mn/Fe] ratio is similar to the nucleosynthesis [Mn/Fe]
ratio of the most massive star that ejects about two orders of
magnitude larger mass of Mn than the lower mass SN II (see Figure 2).

\section{CONCLUSION}

We conclude that each extremely metal-poor star with [Fe/H] $< -2.5$
was formed from a single supernova event and in this early stage the
age-metallicity relation does not hold, but the time elapses toward
lower metallicity. Thus the observed abundance pattern of these stars
which reflect the first generation supernova ejecta enables us to
determine the heavy-element yields ejected from SNe II for any
observed element as a function of the initial stellar mass. It is
noted that we determine the yields descending from the metal-free
progenitor stars. However the metallicity effect on the yields cannot
be seen in the present study such as (i) the Fe yield consistent with
that derived from SN light curve analyses and (ii) the observed [C/Mg]
trend that can be reproduced by the SN model in which the initial
metallicity is solar.

The accuracy of our determination is guaranteed by the quite good
agreement of the IMF-weight ed abundance ratios with the observed
plateau seen in the halo stars for $-2<$ [Fe/H] $<-1$. The yields thus
obtained are indispensable because theoretical SN II models to date
can reliably predict the amounts of only a few elements. Progress in
observations of not only the abundance pattern in extremely metal-poor
stars including the oxygen abundance but also the constant abundance
ratios in metal-poor halo stars promises the acquisition of more
reliable yields for SNe.

As the most intriguing example, we have demonstrated that Eu is
ejected more in less massive stars. This mass dependence is reverse of
those for iron-group elements. We predict that such a difference in
the mass dependence of the ejected mass is (at least partly)
responsible for the difference in the scatter of the observed
abundance ratios.

Combining the yields obtained by our procedure with the Galactic
chemical evolution model, we have shown that the observed abundance
patterns over the whole metallicity range can be well reproduced,
where we introduce two different enrichment phases with different time
scales, e.g., the sequence of the first SN II explosions within
$\sim2\times 10^7$ yr and the well-mixed phase over several Gyrs. The
transition between them will be discussed in the forthcoming paper.

\bigskip

This work has been partially supported by COE research (07CE2002) of
the Ministry of Education, Science, Culture, and Sports in Japan.

\end{document}